\documentclass[a4paper,11pt]{article}

\usepackage{jheppub_mod} 
\usepackage[T1]{fontenc} 

\usepackage{amsmath,booktabs}
\usepackage{graphicx}
\usepackage{xspace}
\usepackage{color}
\pagecolor{white}
\usepackage[dvipsnames]{xcolor}
\usepackage{url}
\usepackage[utf8]{inputenc}
\usepackage{epstopdf}
\usepackage{hyperref}
\usepackage{multirow}
\usepackage{gensymb}
\usepackage{dsfont}
\usepackage{amsfonts, amssymb, xparse, physics, verbatim, tikz, tikz-feynman, slashed, feynmp-auto, caption, booktabs, float, hyperref, enumitem, slashed, verbatim}

\usepackage{ifmtarg}

\newcommand{\mcmule}{{\sc McMule}}

\newcommand{\beq}{\begin{equation}}
\newcommand{\eeq}{\end{equation}}

\newcommand{\Order}{\mathcal{O}}

\newcommand{\GeV}{\,\text{GeV}}
\newcommand{\MeV}{\,\text{MeV}}

\renewcommand{\Im}{\text{Im}\,}
\renewcommand{\Re}{\text{Re}\,}

\newcommand{\A}{\mathcal{A}}


\allowdisplaybreaks[1]

\title{Towards testing \texorpdfstring{$\boldsymbol{(g-2)_\tau}$}{} in \texorpdfstring{$\boldsymbol{e^+e^-\to\tau^+\tau^-}$}{}:\\ radiative corrections and projections for Belle~II}

\author[a]{Jo\"el Gogniat,}
\author[a]{Martin Hoferichter,}
\author[b]{and Yannick Ulrich}

\affiliation[a]{Albert Einstein Center for Fundamental Physics, Institute for Theoretical Physics, University of Bern, Sidlerstrasse 5, 3012 Bern, Switzerland}

\affiliation[b]{Department of Mathematical Sciences, University of Liverpool, Liverpool, L69 3BX, United Kingdom}

\emailAdd{gogniat@itp.unibe.ch}
\emailAdd{hoferichter@itp.unibe.ch}
\emailAdd{yannick.ulrich@liverpool.ac.uk}

\abstract{The arguably most promising avenue towards testing physics beyond the Standard Model in the anomalous magnetic moment of the $\tau$ proceeds via suitably constructed asymmetries in $e^+e^-\to\tau^+\tau^-$ in the presence of a polarized electron beam. 
Such a program, as could be realized at Belle~II assuming a polarization upgrade of the SuperKEKB $e^+e^-$ collider, crucially relies on a careful   
 consideration of radiative corrections. In this work, we present the complete one-loop result for the fully polarized $e^+e^-\to\tau^+\tau^-$ process and its implementation in the Monte-Carlo integrator \mcmule. As an application, we discuss projections relevant for measurements at Belle~II, both with and without electron polarization, and outline
the necessary steps for a generalization to next-to-next-to-leading order. }

\begin{document}
\maketitle
	
\section{Introduction}
\label{sec:intro}

Precision tests of lepton anomalous magnetic moments, $a_\ell=(g-2)_\ell/2$, $\ell=e,\mu,\tau$, have long been key search channels for physics beyond the Standard Model (BSM), reaching back to Schwinger's prediction $a_\ell=\alpha/(2\pi)\simeq 1.16\times 10^{-3}$~\cite{Schwinger:1948iu} and its verification in experiment~\cite{Kusch:1948mvb}.
In the meantime, the exploration of $a_e$ has been pursued at a level below $10^{-12}$, being limited by conflicting input on the fine-structure constant $\alpha$ from atom interferometry experiments  in Cs~\cite{Parker:2018vye} or Rb~\cite{Morel:2020dww}. From the theory side, after a tension in the $5$-loop QED coefficient has  been resolved recently~\cite{Aoyama:2019ryr,Volkov:2019phy,Volkov:2024yzc,Aoyama:2024aly}, the precision is now limited by hadronic vacuum polarization (HVP) to $\simeq 3\times 10^{-14}$~\cite{DiLuzio:2024sps}, while the direct measurement of $a_e$ has reached an uncertainty of $1.3\times 10^{-13}$~\cite{Fan:2022eto}. Given the breadth of ongoing efforts, probing BSM effects in $a_e$ at a level $10^{-13}$ and beyond is certainly realistic.

Hadronic effects become much more critical for $a_\mu$, in which case the experimental determination~\cite{Muong-2:2023cdq,Muong-2:2024hpx}, currently at $2.2\times 10^{-10}$, is soon to improve further with the final release from the Fermilab experiment. While the 2020 report by the Muon $g-2$ Theory Initiative~\cite{Aoyama:2020ynm} (mainly based on Refs.~\cite{Aoyama:2012wk,Aoyama:2019ryr,Czarnecki:2002nt,Gnendiger:2013pva,Davier:2017zfy,Keshavarzi:2018mgv,Colangelo:2018mtw,Hoferichter:2019gzf,Davier:2019can,Keshavarzi:2019abf,Hoid:2020xjs,Kurz:2014wya,Melnikov:2003xd,Colangelo:2014dfa,Colangelo:2014pva,Colangelo:2015ama,Masjuan:2017tvw,Colangelo:2017qdm,Colangelo:2017fiz,Hoferichter:2018dmo,Hoferichter:2018kwz,Gerardin:2019vio,Bijnens:2019ghy,Colangelo:2019lpu,Colangelo:2019uex,Blum:2019ugy,Colangelo:2014qya}) reported a precision of $4.3\times 10^{-10}$ in the SM prediction, subsequently, conflicting new data on the crucial $\pi^+\pi^-$ channel from CMD-3~\cite{CMD-3:2023alj,CMD-3:2023rfe} and lattice-QCD calculations~\cite{Borsanyi:2020mff,Boccaletti:2024guq,RBC:2024fic,Djukanovic:2024cmq,FermilabLatticeHPQCD:2024ppc} have led to tensions that remain to be resolved, e.g., by scrutinizing the role of radiative corrections~\cite{Campanario:2019mjh,Ignatov:2022iou,Colangelo:2022lzg,Monnard:2021pvm,Abbiendi:2022liz,BaBar:2023xiy,Budassi:2024whw,Aliberti:2024fpq}, exploring correlations with other observables~\cite{Crivellin:2020zul,Keshavarzi:2020bfy,Malaescu:2020zuc,Colangelo:2020lcg,Colangelo:2022vok,Colangelo:2022prz,Hoferichter:2023sli,Stoffer:2023gba,Leplumey:2025kvv,Hoferichter:2025lcz}, or independent input on HVP from the MUonE experiment~\cite{CarloniCalame:2015obs,MUonE:2016hru,MUonE:LoI,Banerjee:2020tdt}.
While the target precision for hadronic light-by-light scattering has already been largely achieved thanks to recent improvements, see, e.g., Refs.~\cite{Bijnens:2021jqo,Chao:2021tvp,Blum:2023vlm,Ludtke:2023hvz,Hoferichter:2024fsj,Ludtke:2024ase,Holz:2024lom,Bijnens:2024jgh,Fodor:2024jyn,Hoferichter:2024bae,Hoferichter:2024vbu,Holz:2024diw,Hoferichter:2025yih},
these efforts should also consolidate the HVP contribution in the next years, and thereby the precision test of $a_\mu$~\cite{Colangelo:2022jxc}. However, even with potential experimental advances, e.g., the High Intensity Muon Beam at PSI~\cite{Aiba:2021bxe}, moving beyond a precision of $10^{-10}$ appears challenging.\footnote{After the announcement of the final results from the Fermilab experiment when this paper was under review, the precision of the world average improved to $1.45\times 10^{-10}$~\cite{Muong-2:2025xyk}, compared to $6.2\times 10^{-10}$ for the Standard-Model prediction of the 2025 report by the Muon $g-2$ Theory Initiative~\cite{Aliberti:2025beg}.}

In general, the sensitivity increases with the lepton mass, rendering $a_\tau$, in principle, an even more compelling BSM probe. However, while the SM prediction is well-established~\cite{Eidelman:2007sb,Eidelman:2016aih,Keshavarzi:2019abf,Hoferichter:2025fea}, the short lifetime makes its experimental determination extremely challenging, a rather unfortunate situation given that the interplay of the precision tests of all $a_\ell$ encodes valuable information about the flavor and chirality structure of potential BSM contributions~\cite{Giudice:2012ms,Crivellin:2018qmi,Crivellin:2021rbq,Athron:2021iuf}. This observation had motivated renewed efforts in the last years to extract improved constraints on $a_\tau$.

Traditionally, limits on $a_\tau$ were extracted from $e^+e^-\to e^+e^-\tau^+\tau^-$, either directly from LEP2~\cite{DELPHI:2003nah} or in a global effective-field-theory (EFT) analysis~\cite{Gonzalez-Sprinberg:2000lzf} of LEP and SLD data~\cite{L3:1998lhr,SLD:1999uov,ALEPH:2000fhd}. Since then, measurements of $a_\tau$ in peripheral Pb--Pb collisions at LHC~\cite{delAguila:1991rm} have become available~\cite{ATLAS:2022ryk,CMS:2022arf,CMS:2024qjo}, but even the latest measurement is still a factor three away from probing the Schwinger term. As argued in Ref.~\cite{Crivellin:2021spu}, a meaningful test of realistic BSM scenarios in $a_\tau$ should target a precision of $10^{-6}$.
Reaching such a precision at LHC appears difficult~\cite{Koksal:2017nmy,Gutierrez-Rodriguez:2019umw,Beresford:2019gww,Dyndal:2020yen,Haisch:2023upo,Shao:2023bga,Beresford:2024dsc,Dittmaier:2025ikh}, motivating the consideration of the leptonic channel $e^+e^-\to\tau^+\tau^-$ at Belle II~\cite{Belle-II:2018jsg}. Measuring an absolute cross section at this level would become equally challenging~\cite{Chen:2018cxt,Tran:2020tsj,Krinner:2021cqs}, leaving effects quadratic in spin vectors to extract constraints on $a_\tau$~\cite{Banerjee:2022sgf,Banerjee:2023qjc}, as is the prevalent strategy for the electric dipole moment (EDM) $d_\tau$~\cite{Belle:2002nla,Belle:2021ybo}. However, improved projections have been obtained by constructing so-called transverse and longitudinal asymmetries~\cite{Bernabeu:2007rr,Bernabeu:2008ii}, as possible when a polarized electron beam is available. The original idea in Ref.~\cite{Bernabeu:2007rr} was to consider $e^+e^-\to\tau^+\tau^-$ at the lower $\Upsilon$ resonances, in which case diagrams other than $s$-channel production would be suppressed.

A first step towards realizing such a program was performed in Ref.~\cite{Crivellin:2021spu}, extending the calculation of form-factor-type diagrams to the required two-loop level. At the same time, it was observed that concentrating on the lower $\Upsilon$ resonances did not suffice, first, because devoting a significant  amount of run time below the $\Upsilon(4S)$ is unrealistic and, second, because even at the $\Upsilon(3S)$ resonance continuum $\tau^+\tau^-$ pairs outnumber resonant ones by a factor $10$ due to the spread in beam energy~\cite{BaBar:2020nlq}. Subsequently, first projections were presented in Refs.~\cite{USBelleIIGroup:2022qro,Aihara:2024zds}, concluding that a precision at a level of $10^{-5}$ should be achievable in case  of a polarization upgrade of the SuperKEKB $e^+e^-$ collider, while going beyond would require further advances both regarding statistics and energy calibration. At the same time, the need for the calculation of radiative corrections, including box diagrams, and their implementation in Monte-Carlo (MC) tools was emphasized. In this work, we present progress in this direction.

We calculated the fully polarized next-to-leading-order (NLO) amplitude for $e^+e^-\to\tau^+\tau^-$ and constructed the corresponding asymmetries in the presence of experimental cuts, see Secs.~\ref{sec:overview} and~\ref{sec:asym}. In particular, we implemented the expression into the MC integrator \mcmule, to be able to assess the radiative corrections under realistic experimental conditions and develop a tool that can be used in experiment. 
In Sec.~\ref{sec:results}, we discuss projections directly relevant for Belle II and show that the impact of the box diagrams at NLO remains small, thanks to a suppression by the electron mass. Finally, we outline the next steps, including the generalization to next-to-next-to-leading order (NNLO), in Sec.~\ref{sec:NNLO}, before concluding in Sec.~\ref{sec:conclusions}.

\section{Overview of the calculation}
\label{sec:overview}

\begin{figure}[t]
\centering
\begin{enumerate}[label=(\alph*)]
\item 
\begin{minipage}{0.9\textwidth}
    \includegraphics[width=0.9\textwidth]{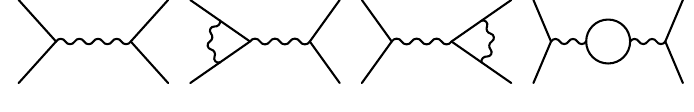}
\end{minipage}
\item 
\begin{minipage}{0.9\textwidth}
    \includegraphics[width=0.44\textwidth]{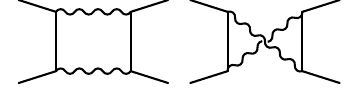}
\end{minipage}
\item 
\begin{minipage}{0.9\textwidth}
    \includegraphics[width=0.9\textwidth]{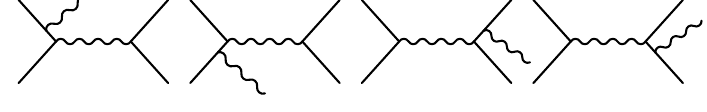}
\end{minipage}
\end{enumerate}
\caption{Leading-order (LO) and NLO Feynman diagrams for $e^+e^- \to \tau^+\tau^-$. (a) Direct $s$-channel diagrams; (b) box diagrams; (c) real emission.}
\label{fig:diagrams}
\end{figure}

In this work we analyze the scattering process
\begin{equation}
    e^+(p_1)e^-(p_2)  \to \tau^+(q_1)\tau^-(q_2)\{\gamma(q_3)\},
\end{equation}
for general kinematics, but having the application to center-of-mass (CM) energies near the $\Upsilon(4S)$ in mind. We incorporate
the fully polarized NLO QED corrections given in Fig.~\ref{fig:diagrams}.\footnote{Diagrams were generated using FeynArts~\cite{Hahn:2000kx} and subsequently evaluated using FeynCalc~\cite{Mertig:1990an} together with Package-X~\cite{Patel:2016fam} through FeynHelpers~\cite{Shtabovenko:2016whf}.} 
For the calculation, we retained the full mass dependence of both electron and $\tau$, along with the polarization vectors for all leptons. Concerning regularization, we employed the four-dimensional formulation of the FDH scheme (FDF)~\cite{Fazio:2014xea} to appropriately handle ultraviolet (UV) and infrared (IR) divergences in $d=4-2\epsilon$ dimensions. Renormalization is then carried out using the on-shell scheme up to NLO QED.
The resulting amplitudes when summed over positron polarization were brought into an analytical form, revealing their symmetry behavior with respect to the scattering angle, see App.~\ref{app:analytic}
and
implemented
in the current stable version {\sc v0.6.0} of the publicly available MC integrator \mcmule~\cite{Banerjee:2020rww, Ulrich:2020frs, Ulrich:2025fij}
\begin{minipage}{\linewidth}
    \hspace{2em}\url{https://mule-tools.gitlab.io}
\end{minipage}
The underlying code is written in Fortran and compiled with \texttt{meson+ninja}. Further details and an accessible beginner walkthrough can be found in the manual\\
\begin{minipage}{\linewidth}
    \hspace{2em}\url{https://mcmule.readthedocs.io}
\end{minipage}
In addition, the \texttt{python} package \texttt{pymule} is provided to streamline both the post-processing and analysis of results. This framework supports computations up to fixed-order NNLO in QED and covers all $2 \to 2$ leptonic processes. 
Once the necessary matrix elements are provided and implemented, for example using OpenLoops~\cite{Buccioni:2017yxi, Buccioni:2019sur}, \mcmule{} handles everything else:
phase space integration is performed using the adaptive MC integrator \texttt{vegas}~\cite{Lepage:1980dq};
the IR subtraction of soft singularities is performed using the $\text{FKS}^\ell$ scheme~\cite{Engel:2019nfw}, an all-order extension of the $\text{FKS}$ scheme~\cite{Frixione:1995ms,Frederix:2009yq};
real-virtual matrix elements are numerically stabilized using next-to-soft (NTS) stabilization~\cite{Banerjee:2021mty,Engel:2021ccn} in delicate regions of phase space.
The user specifies the cuts they want to apply and the histograms they want to compute in an external C++ or Fortran module that is loaded at runtime by \mcmule{}.
This allows for a large amount of flexibility as these observables can be arbitrarily complicated as long as they are IR safe.
For processes that support polarization, this is also where the user can set (or integrate over) polarization vectors of external particles.

Even though \mcmule{} already supports $ee\to\ell\ell$ at NNLO~\cite{Broggio:2022htr,Aliberti:2024fpq}, the implementation mostly (but not completely, \cite{Kollatzsch:2022bqa}) assumes unpolarized particles.
We have extended the calculation to support polarized initial and final states at NLO.
The ingredients necessary for doing the same at NNLO will be discussed in Sec.~\ref{sec:NNLO}.

Using \mcmule, we then computed the total cross section $\sigma_\text{tot}$, as well as the three asymmetries $A_N^\pm, A_T^\pm$, and $A_L^\pm$, see Sec.~\ref{sec:asym}, including cuts tailored to Belle~II kinematics. As input parameters for the numerical calculations we used
\begin{align}
    \alpha &= \frac{e^2}{4\pi} = \frac{1}{137.035999084}, \qquad
    m_e = 0.510998950 \MeV, \qquad 
    m_\tau = 1\,776.86 \MeV, \nonumber\\
    s &= M_{\Upsilon(4S)}^2 = (10\,583.0052 \MeV)^2.
\end{align}
To project out the spins, we defined the operator
\begin{equation}
    P_s = \frac{1}{2}(1+\gamma_5 \slashed{s}) \label{eq:spinprojector}
\end{equation}
and employed the modified completeness relation for the spinors, i.e.,
\begin{equation}
    u(p_i)\Bar{u}(p_i) = \frac{1}{2}(1+\gamma_5\slashed{s}_i)(\slashed{p}_i+m_i),
\end{equation}
where the spin four-vectors $s_i$ satisfy
\begin{equation}
    s_i \cdot p_i = 0, \qquad s_i^2 = -1.
\end{equation}
For the electrons, we assumed polarization along the beam direction so that their polarization vector is given by
\begin{equation}
    s^\mu = \lambda\big(\gamma_e\beta_e, \gamma_e\boldsymbol{\hat{\beta}_e}\big),\qquad \gamma_e = \frac{E_e}{m_e},\qquad \beta_e = \sqrt{1-\frac{1}{\gamma_e^2}},
\end{equation}
with $E_e$, $m_e$ being the energy and mass and $\boldsymbol{\hat{\beta}_e}$ indicating the direction of motion. The helicity $\lambda$ is chosen as $\lambda=\pm 1$. In contrast, to be able to construct the asymmetries, the $\tau$ leptons are polarized along an arbitrary direction. In the $\tau$ rest frame their spin vector is expressed as~\cite{Tsai:1971vv}
\begin{equation}
    s^\mu = (0, s_x, s_y, s_z), \qquad  s_x^2+s_y^2+s_z^2 = 1. 
\end{equation}
Modifying these definitions to take into consideration the case where the $\tau$ leptons are boosted in an arbitrary direction, one obtains for the $\tau^-$ and $\tau^+$ the following vectors
\begin{align}
    s_\mp^\mu &= \begin{pmatrix}
                \pm\gamma_\tau \beta_\tau s_z^\mp \\
                \cos\phi \big(\gamma_\tau s_z^\mp \sin\theta - s_x^\mp \cos\theta\big) + s_y^\mp \sin\phi \\
                \sin\phi \big(\gamma_\tau s_z^\mp \sin\theta - s_x^\mp \cos\theta\big) - s_y^\mp \cos\phi \\
                s_x^\mp \sin\theta + \gamma_\tau s_z^\mp \cos\theta
                \end{pmatrix},
\end{align}
with the azimuthal $\phi$ and polar $\theta$ angles of $\tau^-$. For this specific case, we define the $z$-component to be along the direction of the motion of the $\tau$ and choose the $y$-axis to be perpendicular to the scattering plane. In the chosen reference frame, the momenta of the incoming and outgoing leptons are given by
\begin{equation}
    p_\mp = \begin{pmatrix}
                E_{e^\mp} \\
                0 \\
                0 \\
                \pm\beta_e E_{e^\mp}
                \end{pmatrix},  \qquad 
    q_\mp = \begin{pmatrix}
                E_{\tau^\mp} \\
                \pm\beta_\tau E_{\tau^\mp}\, \cos\phi \sin \theta \\
                \pm\beta_\tau E_{\tau^\mp}\, \sin\phi \sin \theta \\
                \pm\beta_\tau E_{\tau^\mp}\, \cos\theta
                \end{pmatrix},
\end{equation}
where $p_\pm$ denotes the electron momenta $p_1, p_2$, while $q_\pm$ stands for the $\tau$ momenta $q_1, q_2$, respectively,
where, in terms of Mandelstam $s=(p_1+p_2)^2=(q_1+q_2)^2$,
\begin{equation}
  E_{e^\mp}=E_{\tau^\mp}=\frac{\sqrt{s}}{2},\qquad \beta_\ell=\sqrt{1-\frac{4m_\ell^2}{s}},\qquad \gamma_\ell=\frac{\sqrt{s}}{2m_\ell}. 
\end{equation}
The CM scattering angle is related to $t=(p_1-q_1)^2=(p_2-q_2)^2$, $u=(p_1-q_2)^2=(p_2-q_1)^2$, $s+t+u=2m_e^2+2m_\tau^2$, via
\begin{equation}
t
=\frac{1}{2}\Big(2m_e^2+2m_\tau^2-s+ s \beta_e\beta_\tau\cos\theta\Big),\qquad 
u
=\frac{1}{2}\Big(2m_e^2+2m_\tau^2-s- s \beta_e\beta_\tau\cos\theta\Big).
\end{equation}

\section{\texorpdfstring{$\boldsymbol{e^+e^- \to \tau^+\tau^-}$}{ee->tau tau} cross section and asymmetries}
\label{sec:asym}

\subsection{Form factor decomposition}

For the QED form factors, we adopt the standard conventions
\begin{align}
    \bra{p_2}j^\mu\ket{p_1} &= e\, \overline{u}(p_2) \bigg[ \gamma^{\mu} F_1(s) + \frac{i \sigma^{\mu\nu} q_{\nu}}{2 m_{\ell}} F_2(s) + \frac{\sigma^{\mu\nu} q_{\nu} \gamma_5}{2 m_{\ell}} F_3(s) \nonumber\\
    & \phantom{= e\, \overline{u}(p') \big[ \gamma^{\mu} F_1(s) + \frac{i \sigma^{\mu\nu} q_{\nu}}{2 m_{\ell}} F_2(s) \,\,} + \left( \gamma^{\mu} - \frac{2 m_{\ell} q^{\mu}}{q^{2}} \right) \gamma_5 F_4(s) \bigg] u(p_1),
\end{align}
where $p_1$ and $p_2$ denote the momenta of the incoming and outgoing lepton, respectively, and $q = p_2-p_1$. The form factors $F_1(s)$ and $F_2(s)$ are recognized as the Dirac and Pauli form factors and reduce to the charge and the magnetic dipole moment $a_\ell$ in the limit $s=q^2=0$. In contrast, the form factor $F_3(s)$ is $CP$-odd and $F_4(s)$ represents a parity-violating contribution. In the same limit they correspond to the EDM $d_\ell$  and the anapole moment, respectively. To summarize
\begin{align}
    F_1(0) = 1, \qquad F_2(0) = a_\ell, \qquad F_3(0) = \frac{2m_\ell}{e} d_\ell, \qquad F_4(0) = \text{anapole moment}.
\end{align}
Within an EFT framework, potential BSM contributions to $a_\ell$ and $d_\ell$ are encoded by higher-dimensional operators. The leading contribution can be written as 
\begin{align} 
    \mathcal{L}_{\rm dipole} = - C_R^{\ell} \overline{\ell} \sigma^{\mu\nu} P_R \ell F_{\mu\nu} + \text{h.c.}, 
\end{align}
where the right-handed projector is defined as $P_R = \frac{\mathds{1} + \gamma_5}{2}$ and $C_R^\ell$ denotes a complex Wilson coefficient. The real part of $C_R^\ell$ augments the magnetic dipole moment, while the imaginary part gives rise to an EDM. More explicitly, one has 
\begin{align} 
    a_\ell^\text{BSM} = -\frac{4 m_\ell}{e}\Re C_R^\ell, \qquad d_\ell^\text{BSM} = -2 \Im C_R^\ell. 
\end{align} 
In general, the form factors can be decomposed into a SM and a BSM part
\begin{equation} 
    F_{2,3}(s) = F_{2,3}^\text{SM}(s) + F_{2,3}^\text{BSM}(s), 
\end{equation}
with the BSM contributions being suppressed by the new-physics scale $\Lambda_\text{BSM}$. Consequently, in the heavy-new physics limit ($\Lambda_\text{BSM}^2 \gg s$) the leading contributions to the real parts of the form factors are dictated by\footnote{Limits quoted for $\Re d_\tau$, $\Im d_\tau$, e.g., in Refs.~\cite{Belle:2002nla,Belle:2021ybo}, refer to $\Re F_3(s)$ and $\Im F_3(s)$, respectively, so that only the former is actually related to an EDM via the EFT argument.}
\begin{align} 
    \Re F_2(s) \simeq \Re F_2^\text{SM}(s)+ a_\ell^\text{BSM}, \qquad \Re F_3(s) \simeq \frac{2 m_\ell}{e} d_\ell^\text{BSM}, \label{eq:constraints}
\end{align} 
neglecting the tiny SM contribution to $d_\ell$~\cite{Pospelov:2013sca,Ghosh:2017uqq}.
Thus, by computing the SM predictions and subtracting them from the experimental measurements, one can directly constrain the BSM contributions to $a_\ell$ and $d_\ell$. In the remainder of this work, we concentrate on the magnetic and electric dipole moments, outlining the methods and observables used to extract these quantities. Since the anapole moment is not of immediate interest, the $F_4(s)$ contributions will be omitted in the subsequent discussions.

\subsection{Dipole moment extraction via asymmetry-based observables}

Taking into account the $\tau$ decay products, we study the process $e^+e^- \to \tau^+\tau^- \to h^+ \bar{\nu}_\tau h^-\nu_\tau$  with $h^\pm = \pi^\pm, \rho^\pm, a_1^\pm, \dots$ denoting the hadronic final states. By applying the narrow-width approximation (NWA), the production of the $\tau$ and its decay factorize, allowing us to only consider the process $e^+e^- \to \tau^+\tau^-$. Retaining the $\tau$ form factors while restricting ourselves to the direct $s$-channel diagrams, the differential cross section can be expressed as follows
\begin{align}
    \frac{d\sigma}{d\Omega} &= \frac{\alpha^2 \beta_\tau}{4 s \beta_e} \bigg[ |F_1|^2\left(2-\beta_\tau^2 + \beta_e^2\beta_\tau^2\cos^2\theta\right) + |F_2|^2 - |F_3|^2\nonumber\\
    &+ (|F_2|^2+|F_3|^2)\gamma_\tau^2(1-\beta_e^2\beta_\tau^2\cos^2\theta) 
    + 4\Re(F_1F_2^*) + \frac{|F_1+F_2|^2}{\gamma_e^2} \bigg],
\end{align}
where the normalization includes the spin average for the initial-state $e^\pm$ and the entire spin sum. 
By considering semileptonic $\tau$ decays, one can in principle disentangle the only two-loop contribution $\Re(F_1F_2^*)$  to the Pauli form factor from the dominant $|F_1|^2$ terms, since these decays allow for the reconstruction of the $\tau$ production plane and direction of flight~\cite{Tsai:1971vv, Kuhn:1993ra}. However, it has been demonstrated that achieving an accuracy on the order of $10^{-5}$ to $10^{-6}$ is unfeasible, primarily due to systematic uncertainties~\cite{Krinner:2021cqs, Chen:2018cxt}. Therefore, instead of focusing on the total cross section, we will turn to the dipole moment extraction via asymmetry-based observables as proposed in Ref.~\cite{Bernabeu:2007rr}, which can be measured once polarized beams become available. With systematic uncertainties largely canceling out in these asymmetries, reaching a precision of $10^{-5}$ becomes a realistic prospect, while reaching a level of $10^{-6}$ would require more statistics as well as a higher precision measurement of the masses $m_\tau$ and $M_{\Upsilon(1S)}$~\cite{ParticleDataGroup:2024cfk}.

To compute these asymmetries, we must project out the helicity of the electron $\lambda$ and the spins $s_\pm$ of the $\tau^\pm$ using the operator defined in Eq.~\eqref{eq:spinprojector}. In what follows, we concentrate solely on the terms that are linear in the $\tau^\pm$ spins, as all other contributions vanish when constructing the asymmetries. Then, in the absence of electron polarization, i.e., $\lambda=0$, the linearly spin-dependent part of the differential cross section takes the form
\begin{align}
\label{sigma_S}
    \frac{d\sigma^S}{d\Omega} = \frac{\alpha^{2}  \beta_\tau}{8 s \beta_e} \bigg[(s_- - s_+)_x \, X_- + (s_- + s_+)_y \, Y_+ + (s_- - s_+)_z \, Z_-\bigg],
\end{align}
where
\begin{align}
    X_- &= \beta_\tau\beta_e^2 \gamma_\tau \sin\theta \cos\theta \left[ \Im (F_3 F_1^*) + \Im (F_3 F_2^*) \right], \quad Y_+ = \beta_\tau^{2}\beta_e^2 \gamma_\tau \cos\theta \sin\theta \, \Im (F_2 F_1^*), \nonumber\\
    Z_- &= -\beta_\tau \left(\beta_e^2\sin^2\theta + \frac{1}{\gamma_e^2}\right) \left[ \Im (F_3 F_1^*) + \gamma_\tau^{2} \, \Im (F_3 F_2^*) \right],
\end{align}
and only the spin sums for $e^\pm$ are included in the normalization in Eq.~\eqref{sigma_S} (as well as the initial-state spin average).
The spin components can be measured via semileptonic decays~\cite{Bernabeu:2007rr}. In particular, one may substitute
\begin{equation}
    \mathbf{s}_\pm = \mp \alpha_\pm (\sin \theta^*_\pm \cos \phi_\pm, \sin\theta^*_\pm \sin\phi_\pm, \cos\theta^*_\pm),
\end{equation}
where $\phi_\pm$ and $\theta^*_\pm$ denote the azimuthal and polar angles, respectively, of the produced hadron $h^\pm$ in the $\tau^\pm$ rest frame, and $\alpha_\pm$ is the so-called polarization analyzer, discussed further in Sec.~\ref{sec:polana}. It is evident that, without electron polarization, one only gains access to the imaginary parts of the form factors. To probe for instance $\Im F_2$, one can define the normal asymmetry $A_N^\pm$ by utilizing the forward--backward (FB) integration 
\begin{equation}
\label{sigma_FB}
    \sigma_\text{FB} = \left[\int_{0}^{1} dz \frac{d\sigma^S}{d\Omega} - \int_{-1}^{0} dz \frac{d\sigma^S}{d\Omega}\right],
\end{equation}
with \( z = \cos \theta\), and then constructing
\begin{equation}
     \sigma^{\pm}_L = \int_{\pi}^{2\pi} d\phi_{\pm} \frac{d\sigma_{\text{FB}}}{d\phi_{\pm}}, \qquad \sigma^{\pm}_R = \int_{0}^{\pi} d\phi^{\pm} \frac{d\sigma_{\text{FB}}}{d\phi^{\pm}}. \label{eq:ANsigmas}
\end{equation}
Eventually, integrating over all remaining angles as usual yields\footnote{This step includes a factor $4/(4\pi)^2$~\cite{Bernabeu:2007rr,Tsai:1971vv}, to account for $\tau$ spins and angular range in the $\tau^\pm\to h^\pm\nu_\tau$ decays.}
\begin{align}
     A_N^\pm &= \frac{\sigma^\pm_L - \sigma^\pm_R}{\sigma_{\text{tot}}} = \pm \alpha_\pm \frac{\alpha^2 \pi \beta_e \beta_\tau^3 \gamma_\tau}{3 s \sigma_\text{tot}} \Im (F_2 F_1^*). \label{eq:AN}
\end{align}

However, to constrain BSM effects in the magnetic and electric dipole moments, we require observables sensitive to $\Re F_2$ and $\Re F_3$. For this purpose, it is essential to incorporate longitudinally polarized electrons, which are expected to become available with the proposed polarization upgrade of the SuperKEKB collider. The linearly spin-dependent differential cross section receives then additional contributions that are proportional to the electron helicity
\begin{align}
\label{sigmaSlambda}
    \frac{d\sigma^{S\lambda}}{d\Omega} &= \frac{\alpha^2 \lambda \beta_\tau}{16 s \beta_e} \bigg[ 
                        (s_- + s_+)_{x} \, X_+ + (s_- - s_+)_{y} \, Y_- + (s_- + s_+)_{z} \, Z_+ 
                    \bigg],
\end{align}
where
\begin{align}
     X_+ &= \frac{\sin \theta}{\gamma_\tau} \left[ |F_1|^2 + (1 + \gamma_\tau^{2}) \, \Re (F_2 F_1^*) + \gamma_\tau^{2} \, |F_2|^2 \right], \qquad Z_+ = \cos \theta \, |F_1 + F_2|^2, \nonumber\\
     Y_- &= -\beta_\tau \gamma_\tau \sin \theta \left[ \Re (F_3 F_1^*) + \Re (F_3 F_2^*) \right], \label{eq:spinterms}
\end{align}
giving access to the real parts of the form factors. Only the spin sum for $e^+$ is included in the normalization in Eq.~\eqref{sigmaSlambda}  (as well as the initial-state spin average).

One can build again the normal asymmetry, i.e., using the $s_y$-dependent term and employing Eq.~\eqref{eq:ANsigmas}, but now replacing $d\sigma_\text{FB} \to d\sigma_\text{pol}^S$. The helicity difference $d\sigma_\text{pol}^S$ is defined as
\begin{equation}
    d\sigma_{\text{pol}}^S = d\sigma^{S\lambda}|_{\lambda=1} - d\sigma^{S\lambda}|_{\lambda=-1}.
\end{equation}
Note that, in contrast to previous definitions in the literature, cf.\ Ref.~\cite{Crivellin:2021spu,Bernabeu:2007rr}, we do not include an explicit factor of $1/2$; this factor is already absorbed into the definition of the spin projector Eq.~\eqref{eq:spinprojector}. Following the same construction as in Eq.~\eqref{eq:AN}, with
\begin{equation}
     \sigma^{\pm}_{L, F_3} = \int_{\pi}^{2\pi} d\phi_{\pm} \frac{d\sigma_{\text{pol}}^S}{d\phi_{\pm}}, \qquad \sigma^{\pm}_{R, F_3} = \int_{0}^{\pi} d\phi^{\pm} \frac{d\sigma_{\text{pol}}^S}{d\phi^{\pm}}, 
\end{equation}
we obtain an observable that can be used to get constraints on the EDM contributions $d_\tau$
\begin{equation}
    A_{N, F_3}^\pm = \frac{\sigma_{L, F_3}^\pm - \sigma_{R, F_3}^\pm}{\sigma_\text{tot}} = \alpha_\pm\frac{\alpha^2 \pi^2 \beta_\tau^2 \gamma_\tau}{4 s \beta_e \sigma_\text{tot}} \left[\Re (F_3 F_1^*) + \Re (F_3 F_2^*) \right].
\end{equation}
Finally, to derive an observable that probes $a_\tau$, we perform the following integrations in the $\phi_\pm$ or $\theta^*_\pm$ variables~\cite{Bernabeu:2007rr}
\begin{align}
    \sigma^{\pm}_{R, \text{pol}} &= \int_{-\frac{\pi}{2}}^{\frac{\pi}{2}} d\phi_{\pm} \, \frac{d\sigma^S_\text{pol}}{d\phi_{\pm}}, &
    \sigma^{\pm}_{L, \text{pol}} &= \int_{\frac{\pi}{2}}^{\frac{3\pi}{2}} d\phi_{\pm} \, \frac{d\sigma^S_\text{pol}}{d\phi_{\pm}}, \nonumber\\
    \sigma^{\pm}_{\text{FB}, R} &= \int_{0}^{1} dz^{*}_\pm \, \frac{d\sigma^S_{\text{FB}, \text{pol}}}{dz^{*}_\pm}, &
    \sigma^{\pm}_{\text{FB}, L} &= \int_{-1}^{0} dz^{*}_\pm \, \frac{d\sigma^S_{\text{FB}, \text{pol}}}{dz^{*}_\pm},
\end{align}
where $d\sigma^S_{\text{FB}, \text{pol}}$ is defined as in Eq.~\eqref{sigma_FB} with $d\sigma^S\to d\sigma^S_\text{pol}$, 
and subsequently build the transverse $A_T^\pm$ and longitudinal $A_L^\pm$ asymmetries by integrating over the remaining angular variables in the usual manner
\begin{align}
    A_T^\pm = \frac{\sigma^\pm_{R, \text{pol}} - \sigma^\pm_{L, \text{pol}}}{\sigma_{\text{tot}}}, \qquad A_L^\pm = \frac{\sigma^\pm_{\text{FB},R} - \sigma^\pm_{\text{FB},L}}{\sigma_{\text{tot}}}.
\end{align}
Lastly, to cancel out the dominant $|F_1|^2$ term while retaining sensitivity to $\Re F_2$, we define
\begin{align}
   A_T^\pm - \frac{\pi}{2\gamma_\tau} A_L^\pm &= \mp \alpha_\pm \frac{\pi^2 \alpha^2 \beta_\tau^3 \gamma_\tau}{4 s \beta_e \sigma_\text{tot}} \left[\Re (F_2 F_1^*) + |F_2|^2\right], \label{eq:obs}
\end{align}
as the new observable. In fact, we can introduce the effective quantity $\Re F_2^\text{eff}$~\cite{Crivellin:2021spu}
\begin{equation}
    \Re F_2^\text{eff} = \mp \frac{4 s \beta_e \sigma_\text{tot}}{\pi^2 \alpha^2 \beta_\tau^3 \gamma_\tau \alpha_\pm}\left(A_T^\pm - \frac{\pi}{2\gamma_\tau} A_L^\pm\right),
\end{equation}
which is directly amenable to an experimental measurement. By evaluating the corrections and utilizing Eq.~\eqref{eq:constraints}, one can then derive constraints on $a_\tau^\text{BSM}$.

\subsection{Polarization analyzer}
\label{sec:polana}

In Ref.~\cite{Bernabeu:1993er}, the polarization analyzer is introduced as follows
\begin{align}
    \frac{d\Gamma}{\Gamma}\big(\tau^\pm(\mathbf{s})\rightarrow h^\pm(\mathbf{p})+\nu_\tau\big) &= \frac{1}{4\pi}\left[1+\alpha_h \frac{2m_\tau}{m_\tau^2 - m_h^2} \mathbf{p} \cdot \mathbf{s} \right]d\Omega_{h^\pm} \nonumber\\
    &\equiv \frac{1}{4\pi}\big[1+\alpha_h\,\hat{\mathbf{p}}\cdot \mathbf{s} \big]d\Omega_{h^\pm}, \label{eq:defpolana}
\end{align}
where $\mathbf{s}$ denotes the spin vector of the $\tau^-$ lepton, and $\mathbf{p}$ represents the momentum vector of the hadron in the rest frame of the $\tau$ (the second equality defines $\hat{\mathbf{p}}$). The value of $\alpha_h$ for a given decay can be read off from the respective differential decay rate. For spin-$0$ particles such as pions, it should simply be $\alpha_\pi=1$~\cite{Alemany:1991ki, Tsai:1971vv, Hagiwara:1989fn}, but conflicting results, $\alpha_\pi = \frac{m_\tau^2 - 2m_h^2}{m_\tau^2 + 2m_h^2}$~\cite{Bernabeu:1993er, Bernabeu:1994wh} and $\alpha_\pi =  \frac{m_\tau^2 - 2m_h^2}{m_\tau^2 + m_h^2}$~\cite{Bernabeu:2008ii}, have been given in the literature. For this reason, we have reevaluated $\alpha_h$ for spin-$0$ and spin-$1$ final states. 

To do so, we computed the process in which a $\tau^\pm$ decays into an (anti)neutrino and a $W^\pm$-boson, with the $W^\pm$ subsequently decaying into a hadron $h^\pm$. When considering the interaction vertex of the $W^\pm$ boson with $h^\pm$, it is essential to distinguish between the different particle properties. Spin-parity $J^P = 0^-,1^+$ hadrons are only sensitive to the axial-vector current, whereas $J^P = 0^+,1^-$ are sensitive to the vector current~\cite{Donoghue:1992dd}. For scalar and pseudoscalar particles, the interaction vertex is then proportional to $\sim F_h p^\mu$, where $p$ is the momentum of the hadron $h^\pm$ and the decay constant $F_h$ ultimately cancels out in the final expressions. Conversely, for vector and axial-vector particles, the interaction is proportional to $\sim F_h \epsilon^\mu$, with $\epsilon$ being the polarization vector.

When summing over both the longitudinal and transverse spin components of the hadron, we find for the polarization analyzer
\begin{align}
    \alpha_h = \left\{ \begin{array}{lcr}
      1,\qquad & \text{spin 0} \\
      \frac{m_\tau^2 - 2m_h^2}{m_\tau^2 + 2m_h^2},\qquad & \text{spin 1}
    \end{array}\right.. \label{eq:polana}
\end{align}
Alternatively, by disentangling the longitudinal $(\lambda=0)$ and transverse $(\lambda=\pm1)$ contributions, one can reach enhanced sensitivity by considering only the longitudinal part. The explicit forms are
\begin{align}
    \alpha_h^{\lambda=0} = \frac{m_\tau^2}{m_\tau^2+2m_h^2}, \qquad \alpha_h^{\lambda= -1} = -\frac{2 m_h^2}{m_\tau^2+2m_h^2}, \qquad \alpha_h^{\lambda=1} = 0.
\end{align}
For illustrative purposes, some numerical examples are
\begin{align}
    \alpha_h^{\lambda=0} = \begin{cases}
                  0.72 & \qquad h^\pm = \rho^\pm \\[0.1cm]
                  0.51 & \qquad h^\pm = a_1^\pm
                  \end{cases}, \qquad
    \alpha_h^{\lambda=-1} = \begin{cases}
                  -0.28 & \qquad h^\pm = \rho^\pm \\[0.1cm]
                  -0.49 & \qquad h^\pm = a_1^\pm
                  \end{cases}.
\end{align}
It is evident that the sum of these contributions reproduces Eq.~\eqref{eq:polana}. The normalized decay width, as defined in Eq.~\eqref{eq:defpolana}, indicates that nature favors decays in which the $\tau$ spin is aligned with the meson momentum; decays with anti-aligned spins are suppressed. One observes that the polarization analyzer exhibits the same behavior, at least in the limit $m_h \ll m_\tau$. Suppose that the $\tau$ spin is oriented along the positive $z$-axis. Since the neutrino is massless, its chirality coincides with its helicity. Hence, owing to the structure of the weak interaction, the neutrino is necessarily left-handed. Consequently, if the neutrino propagates in the negative $z$-direction, its spin will point in the positive $z$-direction. To conserve total angular momentum at tree level, the hadron must therefore possess a spin projection $\lambda = 0$. This is reflected in the dominant longitudinal component of the polarization analyzer, reaffirming the preference for spin-momentum alignment. Conversely, if the $\tau$ spin is oriented along the negative $z$-direction, the meson must assume a spin projection $\lambda = -1$,  which indeed corresponds to the suppressed contribution to $\alpha_h$. Lastly, the configuration with $\lambda = 1$ vanishes entirely, as it would violate total angular momentum conservation.

\subsection[\texorpdfstring{$\Re F_2$}{F2}-sensitive observable in presence of kinematical cuts]{\texorpdfstring{$\boldsymbol{\Re F_2}$}{F2}-sensitive observable in presence of kinematical cuts}
\label{sec:newcoeff}

The observable defined in Eq.~\eqref{eq:obs}, which gives access to $a_\tau$, is strongly affected by the chosen kinematical cuts. This is because $A_T^\pm$ and $A_L^\pm$ are modified to a different extent, leading to residual contributions dependent on $|F_1|^2$. It is important to notice that this observable also receives contributions from the $\gamma $--$ \Upsilon $--$ Z$ interference. Although such corrections have been shown to be necessary to include, these interference terms exhibit the same angular distribution as the $|F_1|^2$ term~\cite{Bernabeu:2007rr}. Hence, the previously defined linear combination also effectively eliminates the $\gamma $--$ \Upsilon $--$ Z$ corrections. Consequently, it is essential to replace the factor $\frac{\pi}{2\gamma_\tau}$ with a function $\mathcal{C}(\Theta)$ that depends on the angular cuts, thereby ensuring that the unwanted contributions are properly canceled even in the presence of these kinematical constraints.

We assume that the angular cuts remain symmetric in the CM frame, an approximation that holds nearly true for soft photons. Specifically, let the angular cuts range from \( \Theta \) to \( \pi - \Theta \). As one concludes from Eq.~\eqref{eq:spinterms}, we require that
\begin{align}
    \int_\Theta^{\pi - \Theta} d\theta\, \frac{\sin^2\theta}{\gamma_\tau} - \mathcal{C}(\Theta)\left(\int_\Theta^{\frac{\pi}{2}} d\theta\, \cos\theta \sin\theta - \int_{\frac{\pi}{2}}^{\pi - \Theta} d\theta\, \cos\theta \sin\theta\right) \overset{!}{=} 0.
\end{align}
Solving this equation is straightforward and yields
\begin{align}
    \mathcal{C}(\Theta) = \frac{\pi/2 - \Theta + \sin\Theta \cos\Theta}{\gamma_\tau \cos^2\Theta}, \label{eq:newcoeff}
\end{align}
which simplifies in the special case of no cuts, i.e., $\Theta = 0$, indeed to
\begin{align}
    \mathcal{C}(0) = \frac{\pi}{2\gamma_\tau}.
\end{align}
Experimentally, the observables $A_T^\pm$ and $A_L^\pm$ are defined using event-count binnings~\cite{USBelleIIGroup:2022qro}. When angular cuts are applied, these binnings must be adjusted accordingly. For instance, in the case of $A_L^\pm$ one would redefine
\begin{enumerate}
    \item $N^\text{FF}_\pm(R_e \text{ or } L_e): \#\, \tau^\pm$ decays with $\phantom{-}0<\cos\theta<\cos\Theta$ and $h^\pm$ in $\phantom{-}0<\cos\theta_\pm^*<1$,
    \item $N^\text{FB}_\pm(R_e \text{ or } L_e): \#\, \tau^\pm$ decays with $\phantom{-}0<\cos\theta<\cos\Theta$ and $h^\pm$ in $-1<\cos\theta_\pm^*<0$,
    \item $N^\text{BF}_\pm(R_e \text{ or } L_e): \#\, \tau^\pm$ decays with $-\cos\Theta<\cos\theta<0$ and $h^\pm$ in $\phantom{-}0<\cos\theta_\pm^*<1$,
    \item $N^\text{BB}_\pm(R_e \text{ or } L_e): \#\, \tau^\pm$ decays with $-\cos\Theta<\cos\theta<0$ and $h^\pm$ in $-1<\cos\theta_\pm^*<0$,
\end{enumerate}
with the scattering angle $\theta$, the hadronic angles $\theta^*_\pm$, and $R_e \text{ or } L_e$ denoting the electron helicity. The same applies for the bins of $A_T^\pm$. Since the cut angle cannot be determined exactly, an inherent uncertainty $\Delta\Theta$ is introduced. Consequently, this systematic uncertainty propagates into the full observable $\mathcal{O} \doteq A_T^\pm - \mathcal{C}(\Theta)A_L^\pm$ as follows
\begin{equation}
    \Delta \mathcal{O} = \left| \frac{d \mathcal{O}}{d \Theta} \right|\Delta\Theta = \left|\frac{\partial A_T^\pm}{\partial \Theta} - \mathcal{C}\frac{\partial A_L^\pm}{\partial \Theta} - \frac{\partial C}{\partial \Theta} A_L^\pm\right| \Delta\Theta, 
\end{equation}
where the derivatives can be approximated computationally. A rough estimate is obtained by considering only the direct $s$-channel diagram contributions to $\mathcal{O}$ while neglecting the box diagrams and bremsstrahlung, since they are the dominant contributions, at least when only soft photons are present. Evaluating $\left| \frac{d \mathcal{O}}{d \Theta} \right|$ under these assumptions yields a magnitude on the order of $\Order(10^{-4})$. Consequently, even an angular variation $\Delta \Theta$ of $1^\circ \simeq 0.017~\text{rad}$ would reduce the uncertainty to a few times $10^{-6}$. A more thorough error estimate is obtained by performing a MC run of $A_T^\pm$ and $A_L^\pm$ with angular cuts set slightly above and below those tailored to Belle~II used in Sec.~\ref{sec:results}. Runs were carried out with $17.1^\circ < \theta^*_{\tau^\pm} < 149.9^\circ$ and $16.9^\circ < \theta^*_{\tau^\pm} < 150.1^\circ$, which yields an estimate for the error propagation of $\left| \frac{d \mathcal{O}}{d \Theta} \right| = 0.0009(2)$ and together with $\Delta\Theta=1^\circ$ would induce an error of $1.6\times 10^{-5}$. To further minimize this uncertainty, it should also be possible to perform the analysis for slightly different values of $\Theta$ and interpolate, to better enforce the cancellation according to Eq.~\eqref{eq:newcoeff}.

A further caveat arises from the present definitions for $A_L^\pm$ and $A_T^\pm$ in Ref.~\cite{USBelleIIGroup:2022qro}, where one normalizes to the sum of events in each bin separately. This relies on the assumption that the unpolarized cross section is symmetric in $\cos\theta$. However, when real corrections are taken into account, even in the case of soft photons, the symmetric behavior is no longer guaranteed. As a result, for $A_L^\pm$ it is only possible to normalize two separate regions rather than four, namely
\begin{align}
    A_F^\pm &\equiv \frac{N^\text{FF}_\pm(R_e)-N^\text{FF}_\pm(L_e)-N^\text{BF}_\pm(R_e)+N^\text{BF}_\pm(L_e)}{N^\text{FF}_\pm(R_e)+N^\text{FF}_\pm(L_e)+N^\text{BF}_\pm(R_e)+N^\text{BF}_\pm(L_e)}, \nonumber\\
    A_B^\pm &\equiv \frac{N^\text{FB}_\pm(R_e)-N^\text{FB}_\pm(L_e)-N^\text{BB}_\pm(R_e)+N^\text{BB}_\pm(L_e)}{N^\text{FB}_\pm(R_e)+N^\text{FB}_\pm(L_e)+N^\text{BB}_\pm(R_e)+N^\text{BB}_\pm(L_e)},
\end{align}
which need to be combined to $A_L^\pm$.
The same considerations apply to the transverse asymmetry.

\section{Results}
\label{sec:results}

We compared our unpolarized and longitudinally polarized matrix element results with earlier computations~\cite{Kollatzsch:2022bqa} as well as with the amplitude generator OpenLoops~\cite{Buccioni:2017yxi, Buccioni:2019sur}. In both comparisons, we observed full agreement. The results in the following concern the process $e^+e^- \to \tau^+\tau^-$ at $s=M_{\Upsilon(4S)}^2$.

Before we included kinematical constraints specified to Belle~II, we further validated our approach by computing the asymmetries \(A_N^-\), \(A_T^-\), and \(A_L^-\) without any angular cuts, and comparing them to the analytical expressions found in Ref.~\cite{Crivellin:2021spu}. Note that in Ref.~\cite{Crivellin:2021spu} the total cross section $\sigma_\text{tot}$ was kept at tree level and only the direct $s$-channel diagrams were considered for the one-loop corrections. We adopted this for the numerical results found in Table~\ref{tab:reswocuts}. Our calculations reproduce the analytical predictions within errors, hence, confirming our method once more. All the raw data, both with and without cuts as well as the analysis codes and plots, are available at\\
\begin{minipage}{\linewidth}
    \url{https://mule-tools.gitlab.io/user-library/dilepton/belleII-tau-g-2}
\end{minipage}

\begin{table}[t]
  \centering
  \renewcommand{\arraystretch}{1.3}
  \begin{tabular}{@{}lcc@{}} 
    \toprule
   Observable & Prediction & Results from \mcmule\\
    \midrule
    $A_N^-$ & $-0.0001366$ &  $-0.0001364(5)$ \\
   $A_T^--\frac{\pi}{2\gamma_\tau}A_L^-$ & $-0.000359$ &  $-0.000359(2)$ \\
    \bottomrule
  \end{tabular}
  \renewcommand{\arraystretch}{1.0}
  \caption{Results obtained from \mcmule{} for the asymmetries without angular cuts compared to theoretical predictions.}
  \label{tab:reswocuts}
\end{table}

\subsection{Projections for Belle~II}

In the following, we focus on the case with angular constraints tailored to Belle~II while including only soft photons. Realistic detector approximations are provided in Ref.~\cite{Belle:2000cnh}, and accordingly the scattering angle and photon energy are chosen to be
\begin{equation}
    17^\circ < \theta_{\tau^\pm}^\text{lab} < 150^\circ, \qquad E_\gamma < 50~\text{MeV}.
\end{equation}
We want to stress that these preliminary cuts are adjustable and can be changed to any other values if required. Those cuts are set in the laboratory (lab) frame, defined by the electron and positron beam energies given as
\begin{equation}
    E_{e^-} = 7\GeV, \qquad E_{e^+} = 4\GeV.
\end{equation}
However, notice that the FB integrations needed for calculating the asymmetries are carried out in the CM frame of the $\tau$ leptons. In this frame, the angular cuts are nearly symmetric, because identical cuts are applied to both $\tau^\pm$ and only soft photons are considered. The constraints are roughly given by
\begin{equation}
    22.66^{\circ} < \theta_{\tau^\pm} < 157.34^{\circ}.
    \label{cut_CM}
\end{equation}
Including hard photons would break this symmetry. Moreover, the asymmetric behavior with respect to the scattering angle of the NLO distributions, clearly seen in Figs.~\ref{fig:totcrossCMS}--\ref{fig:ALCMS} found in App.~\ref{app:CMSresults}, arises from the bremsstrahlung diagrams. In our scenario, boosting from the CM to the lab frame shifts the angular distributions to smaller angles. Specifically, for forward emission ($0^\circ < \theta_{\tau^\pm} < 90^\circ$), the distributions will be compressed, while for backward emission ($90^\circ < \theta_{\tau^\pm} < 180^\circ$) the distribution is effectively stretched out.

\begin{figure}[t]
    \centering
    \includegraphics[width=0.8\linewidth]{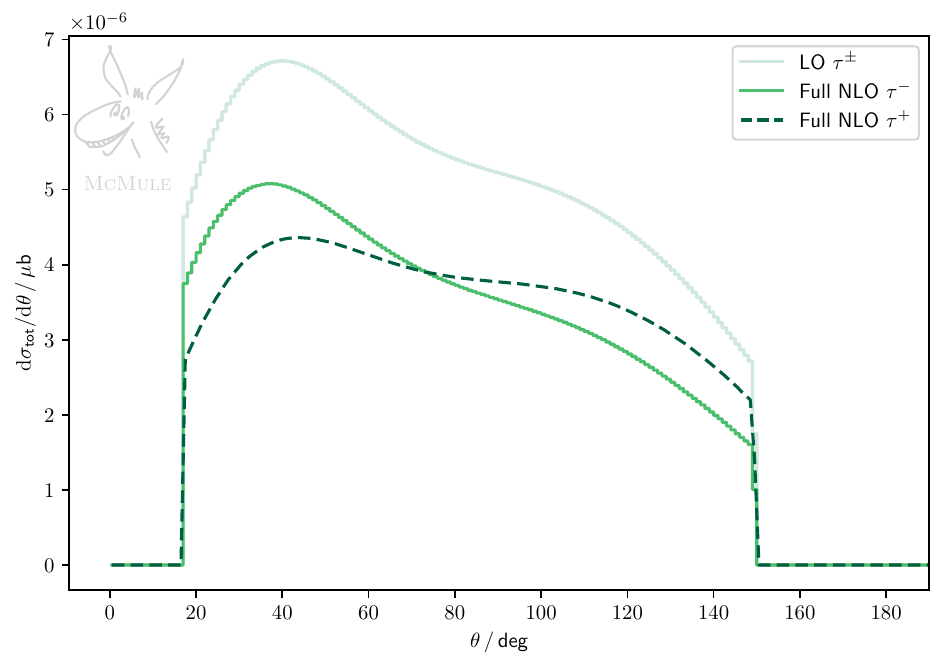}
    \caption{Numerical results of the differential cross section obtained with \mcmule{} including cuts.}
    \label{fig:totcrossLab}
\end{figure}

\begin{figure}[t]
    \centering
    \includegraphics[width=0.8\linewidth]{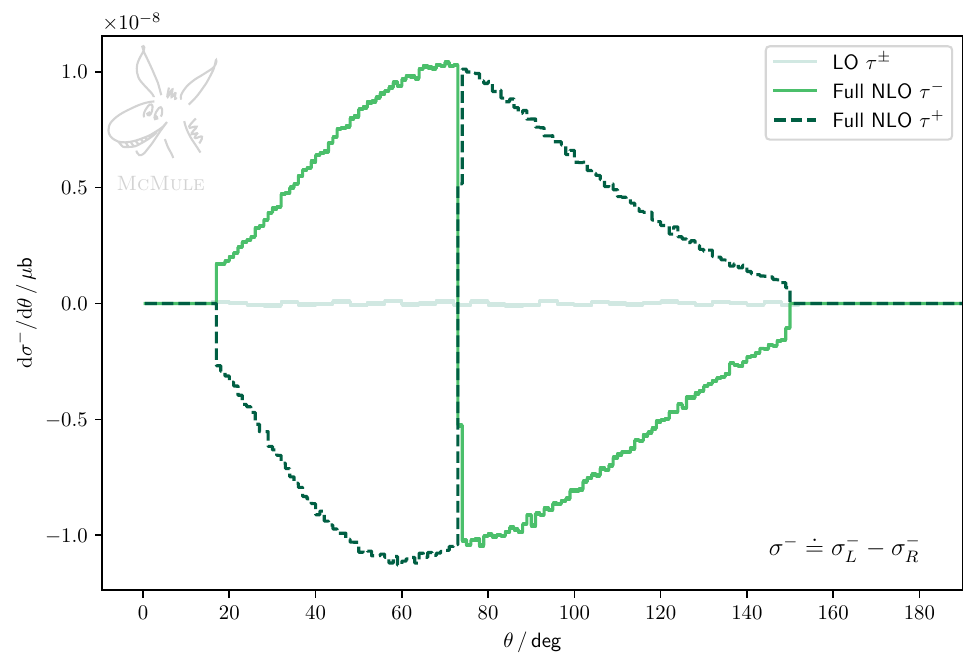}
    \caption{Numerical results of the differential angular distribution of $\A_N^- \times \sigma_\text{tot}$ obtained with \mcmule{} including cuts.}
    \label{fig:ANLab}
\end{figure}

\begin{figure}[t]
    \centering
    \includegraphics[width=0.8\linewidth]{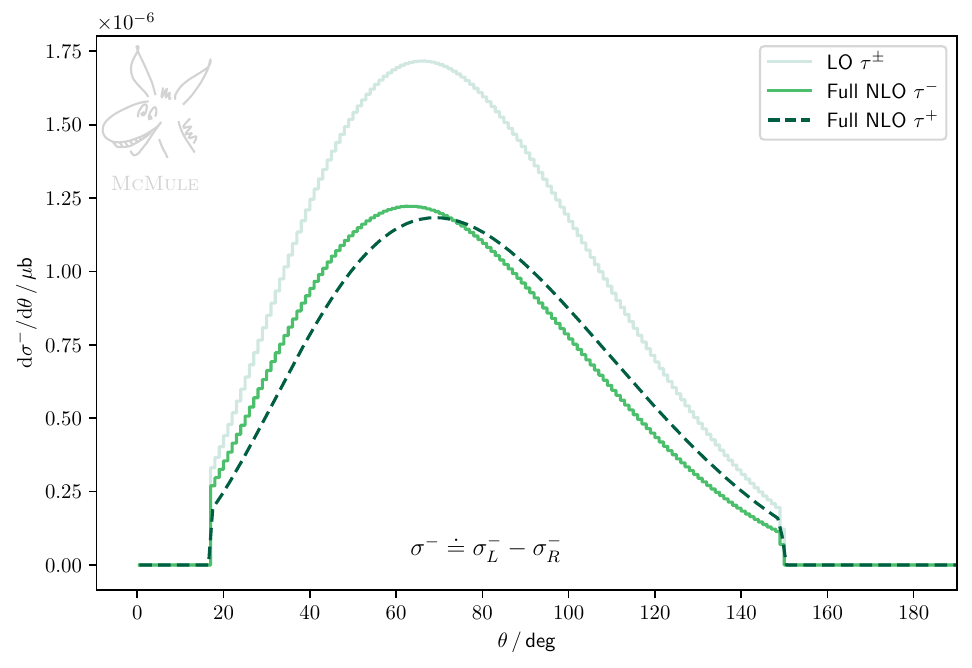}
    \caption{Numerical results of the differential angular distribution of $\A_T^- \times \sigma_\text{tot}$ obtained with \mcmule{} including cuts.}
    \label{fig:ATLab}
\end{figure}

\begin{figure}[t]
    \centering
    \includegraphics[width=0.8\linewidth]{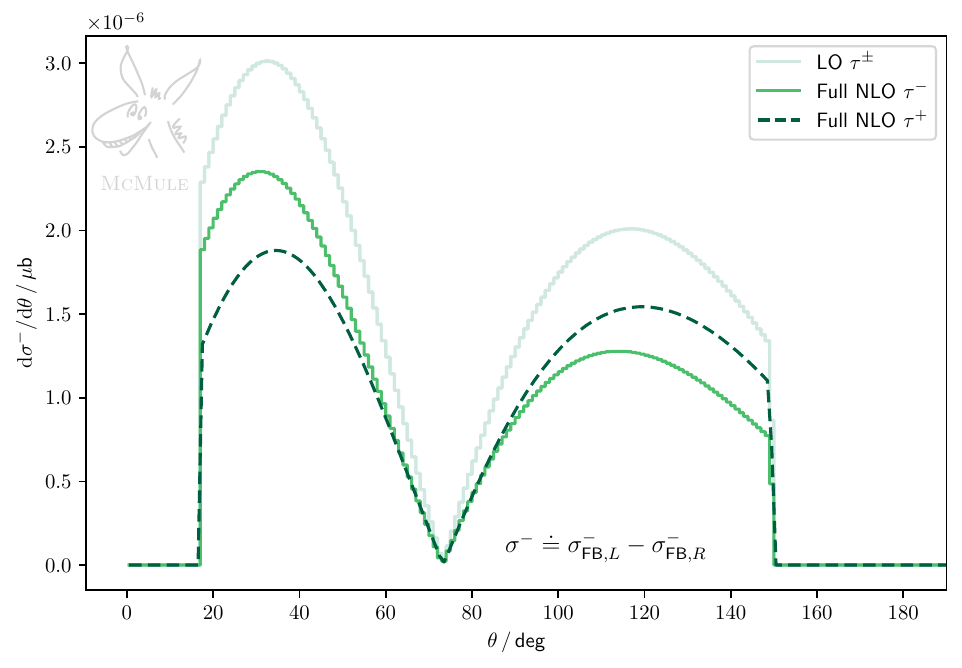}
    \caption{Numerical results of the differential angular distribution of $\A_L^- \times \sigma_\text{tot}$ obtained with \mcmule{} including cuts.}
    \label{fig:ALLab}
\end{figure}

To compute the asymmetries $A_N^-, A_T^-$, and $A_L^-$ using \mcmule, we first evaluated the denominator of the asymmetries, i.e., the total cross section $\sigma_\text{tot}$, and then the numerators separately. The result of the differential cross section in the lab frame is given in Fig.~\ref{fig:totcrossLab}, 
while the differential angular distributions for the numerators in the lab frame are presented in Figs.~\ref{fig:ANLab}--\ref{fig:ALLab}. 
The distributions are functions of the scattering angle between the electron and $\tau^-$ (or $\tau^+$). By integrating the distributions in Figs.~\ref{fig:ANLab}--\ref{fig:ALLab} as well as normalizing by $\sigma_\text{tot}$, we directly obtain the desired asymmetries. The resulting numerical values are summarized in Table~\ref{tab:reswithcuts}. 
\begin{table}[t]
  \centering
  \renewcommand{\arraystretch}{1.3}
  \begin{tabular}{@{}lc@{}} 
    \toprule
    Observable & Results from \mcmule\\
    \midrule
    $A_N^-$ &  $-0.0002061(8)$ \\
    $A_T^- -\frac{\pi}{2\gamma_\tau}A_L^-$ &  $\phantom{-}0.0253360(8)$ \\
    \bottomrule
  \end{tabular}
  \renewcommand{\arraystretch}{1.0}
  \caption{Results obtained from \mcmule{} for the asymmetries with angular cuts at NLO including box diagrams.}
  \label{tab:reswithcuts}
\end{table}

The differences between the results in Tables~\ref{tab:reswocuts} and~\ref{tab:reswithcuts} arise not only from the application of cuts but also from evaluating $\sigma_\text{tot}$ at NLO rather than at tree level and including real corrections. Although box diagrams were incorporated in Table~\ref{tab:reswithcuts}, their impact appears to be negligible, as discussed in Sec.~\ref{sec:box}. Furthermore,  the value of the $\Re F_2$-sensitive observable appears to change its sign and is two orders of magnitude larger than expected, given that both the $\Im F_2$ and $\Re F_2$-sensitive observables should be of the order $\order{10^{-4}}$. This discrepancy arises because the angular cuts affect the longitudinal asymmetry much more significantly than the transverse asymmetry, clearly displayed in Figs.~\ref{fig:ALLab} and~\ref{fig:ATLab}, respectively. Due to this disparity, remnants of the dominant $|F_1|^2$ contributions, which canceled out when using no cuts, reappear again. This issue is already discussed and resolved in Sec.~\ref{sec:newcoeff}. Consequently, the original coefficient $\frac{\pi}{2 \gamma_\tau}$ must be replaced with $\mathcal{C}(\Theta)$, where $\Theta = 22.6597^\circ$. Hence, making use of Eq.~\eqref{eq:newcoeff}, we obtain
\begin{equation}
    \mathcal{C}(\Theta) = 0.603638.
\end{equation}
By employing the new coefficient, we finally derive a result that effectively cancels both the dominant charge form factor corrections as well as the $\gamma $--$ \Upsilon $--$ Z$ interference contributions even when applying cuts. The corresponding new results can be found in Table~\ref{tab:reswithnewcoeff}.

\begin{table}[t]
  \centering
  \renewcommand{\arraystretch}{1.3}
  \begin{tabular}{@{}lc@{}} 
    \toprule
    Observable & Results from \mcmule\\
    \midrule
    $A_N^-$ &  $-0.0002061(8)$ \\
    $A_T^- -\mathcal{C}(\Theta)A_L^-$ &  $-0.0004256(9)$ \\
    $A_T^- -\mathcal{C}(\Theta)A_L^- \big|_\text{LO}$ &  $\phantom{-}0.0000001(3)$ \\
    \bottomrule
  \end{tabular}
  \renewcommand{\arraystretch}{1.0}
  \caption{Final results obtained from \mcmule{} for the asymmetries with cuts at NLO including box diagrams. Here, the new coefficient $\mathcal{C}(\Theta)$ was used instead of $\frac{\pi}{2 \gamma_\tau}$.}
  \label{tab:reswithnewcoeff}
\end{table}

In particular, these results show the right sign ($-$) and the correct order of magnitude $(\simeq 10^{-4})$ for the $\Re F_2$-sensitive observable. As an additional verification of the form of the new coefficient $\mathcal{C}(\Theta)$, note that $A_T^--\mathcal{C}(\Theta)A_L^- \big|_\text{LO}$ evaluated at tree level yields zero (within errors). This is precisely the cancellation required to eliminate the unwanted contributions mentioned earlier. Note that the results presented here do not take the polarization analyzer into account, see Sec.~\ref{sec:polana}. Accordingly, when considering the contributions from a specific hadron $h^\pm$, the numerical values obtained with \mcmule{} must be multiplied by the factor $\alpha_h$ as defined in Eq.~\eqref{eq:polana}.

\subsection{Impact of box diagrams}
\label{sec:box}

In addition to the direct results for the observables, we were also interested in the impact of the box diagrams. To this end, it is instructive to examine the symmetry behavior of the contributions collected in App.~\ref{app:analytic} under the transformation $\cos\theta \leftrightarrow -\cos\theta$ or equivalently the exchange of the Mandelstam variables $t \leftrightarrow u$.
Only the contributions that mirror the behavior of the $s$-channel diagrams will influence the asymmetries. As summarized in Table~\ref{fig:symmetry}, all dominant contributions from the box diagrams ultimately vanish. Only a negligible $\order{m_e^2}$-contribution persists for the observable sensitive to $\Re F_2$, which is of the order of $\simeq 10^{-10}$, compared to the dominant contribution of approximately $\simeq 10^{-4}$. For this reason, the one-loop box diagrams can be considered almost negligible, at least within our current precision goals. This conclusion still persists in the presence of the angular cut~\eqref{cut_CM}, because the symmetry remains preserved in the CM frame. A similar suppression does not arise for the real contributions.

\begin{table}[t]
  \centering
  \renewcommand{\arraystretch}{1.3}
  \begin{tabular}{@{}lcc@{}}
    \toprule
    Spin & $s$-channel diagrams & Box diagrams \\
    \midrule
    $s_x$ & $\text{even}$ &  $\text{odd} + \order{m_e^2}\,\, \text{even}$ \\
    $s_y$ & $\text{odd}$ &  $\text{even}$ \\
    $s_z$ & $\text{odd}$ &  $\text{even} + \order{m_e^2}\,\, \text{odd}$ \\
    Independent & $\text{even}$ &  $\text{odd}$ \\
    \bottomrule
  \end{tabular}
  \renewcommand{\arraystretch}{1.0}
  \caption{Symmetry behavior of the one-loop contributions under $t\leftrightarrow u$.}
  \label{fig:symmetry}
\end{table}

\section{Next steps: beyond NLO and NWA}
\label{sec:NNLO}

Eventually this calculation will require NNLO accuracy.
For the unpolarized case, the full NNLO corrections have been addressed for $e\mu\to e\mu$ in Ref.~\cite{Broggio:2022htr} and for $ee\to\mu\mu$ in Ref.~\cite{Aliberti:2024fpq} using \mcmule.
This calculation relied on massification~\cite{Engel:2018fsb} for the two-loop amplitude and NTS stabilization~\cite{Banerjee:2021mty} and OpenLoops~\cite{Buccioni:2017yxi, Buccioni:2019sur} for the real-virtual corrections.
Massification allows us to recover all terms in $m_e$ that are not power-suppressed from the calculation with $m_e=0$~\cite{Bonciani:2021okt}.

To achieve NNLO accuracy, the two-loop calculation of Ref.~\cite{Bonciani:2021okt} would have to be repeated with a full form factor decomposition of the $\tau$ and electron.
While technically non-trivial, this does not increase the complexity of the integration-by-part reduction or master integrals required and should be relatively straightforward.
However, because the calculation is carried out with $m_e=0$, only two out of the four possible helicity configurations for the initial state can be recovered.
The remaining two configurations are power-suppressed,  and thus not needed for this application.

Next, the public version of OpenLoops only supports polarization along the momentum direction.
This needs to be extended to support arbitrary polarization directions.
Luckily, the structure of OpenLoops makes this fairly straightforward and work is ongoing to implement this in the public release~\cite{maxprivcom}.

Finally, the NTS stabilization procedure that is being used to improve speed and convergence of the integration needs to support polarization.
This was first developed in Ref.~\cite{Kollatzsch:2022bqa} and it is fairly trivial to combine it with the procedure proposed in Refs.~\cite{Bonocore:2021cbv,Balsach:2023ema}.
However, as discussed in Ref.~\cite{Engel:2023ifn} an extra piece that corresponds to the insertion of a magnetic operator needs to be considered.
Since the Wilson coefficient of this operator is already $\mathcal{O}(\alpha)$ this calculation is only needed at tree-level and it is relatively straightforward.
Needless to say that thorough testing and validation will be required.

Moreover, the treatment of $\tau$ decays requires further investigation.
At NNLO precision, corrections beyond the NWA become potentially important and must be explored in the near future.
This involves long-range interactions between the electron and $\tau$ decay products.
The equivalent cases have already been considered in QCD, e.g., in the context of Drell--Yan~\cite{Dittmaier:2015bfe,Dittmaier:2014qza} and it should be possible to extend this work to our case.

Lastly, the current MC generator used by Belle~II, \texttt{TAUOLA}~\cite{Jadach:1993hs, Chrzaszcz:2016fte}, which has been extensively used to simulate $\tau$ decays, needs to be significantly improved to meet the high-precision requirements, e.g., for hadronic decays one avenue of improvement concerns structure-dependent corrections.

\section{Conclusions}
\label{sec:conclusions}

In this work, we evaluated the NLO radiative corrections to $e^+e^-\to\tau^+\tau^-$, as required for testing the anomalous magnetic moment of the $\tau$ using asymmetries that become accessible with a polarized electron beam.  
To this end, we computed the complete one-loop QED corrections for the fully polarized process in analytic form. Furthermore, we performed a numerical evaluation and presented results for both the total cross section $\sigma_\text{tot}$, as well as the asymmetries $A_N^\pm$, $A_T^\pm$, and $A_L^\pm$ for Belle~II kinematics, implemented in the MC integrator \mcmule{}.

A significant aspect of our study concerns the investigation of the impact of realistic kinematical cuts, specifically adapted to Belle~II detector conditions, on these asymmetries. We demonstrated that while the dominant $|F_1|^2$ contributions cancel in the ideal case, angular cuts introduce residual effects. To mitigate these, we proposed an alternative linear combination of $A_T^\pm$ and $A_L^\pm$ that uses a modified coefficient $\mathcal{C}(\Theta)$ to effectively cancel both the dominant charge form factor corrections as well as the $\gamma$--$\Upsilon$--$Z$ interference. Notably, this new combination is less sensitive to potential uncertainties in the angular cuts, thereby keeping systematic uncertainties under control. Our numerical analysis confirms that this prescription restores the expected order of magnitude and sign for the $\Re F_2$-sensitive observable. We further reexamined the polarization analyzer, which quantifies the prefactor in the reconstruction of the $\tau$ spin information via semileptonic decays, demonstrating that in the case of spin-$1$ final states the sensitivity of the measurement can be significantly enhanced if the longitudinally polarized component can be isolated. 

In addition, we thoroughly investigated the impact of box diagrams on our observables. Our analysis reveals that the leading contributions of the box diagrams exhibit a symmetry behavior opposite to that of the direct $s$-channel diagrams, resulting in a cancellation of the box contributions for the asymmetries. In fact, only subleading corrections of $\Order(m_e^2)$, with a magnitude of approximately $10^{-10}$, persist in the observable relevant for constraining $a_\tau$. This degree of suppression confirms that, within our precision goals, the box diagrams are effectively negligible. 

From these studies, we see no fundamental issues in reaching a precision of $10^{-5}$ and beyond, essential for establishing relevant bounds on potential BSM contributions. However, to reach this goal, the calculation should be extended to NNLO, to ensure that the theoretical precision does not become the limiting factor. 
We outlined the essential steps required to reach this objective, with key challenges including the recalculation of two-loop amplitudes incorporating full polarization, implementation and validation of the NTS stabilization procedure, and improved treatments of $\tau$ decays beyond the NWA. Work along these lines is in progress.

\acknowledgments

We thank J.~Michael Roney for valuable discussions.
We would further like to thank William J. Torres Bobadilla for discussions regarding the recalculation of the two-loop amplitude and Max Zoller regarding the possibility to implement more general polarization in OpenLoops.
Finally, we would like to thank the rest of the \mcmule{} Collaboration and especially Sophie Kollatzsch and Tim Engel for discussions regarding implementation.
Financial support by the SNSF (Project No.\ TMCG-2\_213690) is gratefully acknowledged.

\appendix

\section{Analytical results of one-loop corrections}
\label{app:analytic}

In the following, the explicit analytical expressions for the one-loop diagrams appearing in Fig.~\ref{fig:diagrams} interfered with the tree-level amplitude are presented. The expressions are also provided as supplemental
material in the form of a \textsc{Mathematica} notebook.

\subsection{Tree level}
\begin{align}
    \mathcal{M}_\text{tree}
    &= 2\alpha^2 \pi^2\bigg[\frac{2}{\gamma_\tau}\lambda s_x \sin\theta + 2\lambda s_z \cos\theta +2-\beta_\tau^2\big(1-\beta_e^2 \cos^2\theta\big)+1-\beta_e^2\bigg]. 
\end{align}

\subsection{Electron vertex correction}

\begin{align}
    \mathcal{M}_{ee} &= \frac{\alpha}{\pi} \mathcal{M}_\text{tree}\left(\frac{\mu^2}{m_e^2}\right)^\epsilon\notag\\
    &\times\bigg[\frac{(4m_e^2-s)+(2m_e^2-s)\Re B_0^{e}(s)}{(s-4m_e^2)\epsilon} -2   +(2m_e^2-s)\Re \bar C_0^{e}(s) -\frac{3}{2}\Re B_0^{e}(s)\bigg]   \nonumber\\
    &+4\alpha^3\pi \frac{m_e^2\left(1-\beta_\tau^2\cos^2\theta\right)}{s}\Re B_0^{e}(s). 
\end{align}

\subsection{Tau vertex correction}

\begin{align}
    \mathcal{M}_{\tau\tau} &= \frac{\alpha}{\pi} \mathcal{M}_\text{tree}\left(\frac{\mu^2}{m_\tau^2}\right)^\epsilon \notag\\
    &\times\bigg[\frac{(4m_\tau^2-s)+(2m_\tau^2-s)\Re B_0^{\tau}(s)}{(s-4m_\tau^2)\epsilon} -2   +(2m_\tau^2-s)\Re \bar C_0^{\tau}(s) -\frac{3}{2}\Re B_0^{\tau}(s)\bigg]   \nonumber\\
    &+4\alpha^3\pi \frac{m_\tau^2\left(1-\beta_e^2\cos^2\theta\right)}{s}\Re B_0^{\tau}(s)\notag\\
    &+\alpha^3\pi\Big\{ s_y \sin(2\theta)\frac{\beta_e^2}{2\gamma_\tau} \Im B_0^{\tau}(s)+  \lambda s_x \sin\theta \frac{1}{\gamma_\tau}\Re B_0^{\tau}(s)\Big\}.
\end{align}

\subsection{Box diagrams}

$s_y$-component:
\begin{align}
   \mathcal{M}_{e\tau}^{(y)} &= -\frac{2\alpha^3\pi s_y \sin\theta\beta_e\beta_\tau}{((m_e^2-m_\tau^2)^2-tu)\gamma_\tau}\Bigg\{-\frac{\pi}{2}s^2\beta_e^2\sin^2\theta+ s^2 m_e^2 \beta_e^2 \Im C_0^{e}(s)\nonumber\\
    &+s^2\left[2m_e^2\beta_\tau^2-m_\tau^2\beta_e^2\sin^2\theta\right] \Im C_0^{\tau}(s) \nonumber\\
    &+sm_e^2(m_e^2-m_\tau^2-3t)\left(\frac{\pi t}{\lambda_{e \tau}(t)}\Re B_0^{e \tau}(t) - \Im \bar C_0^{e \tau}(t)\right) \nonumber\\
    &+sm_e^2(m_e^2-m_\tau^2-3u)\left(\frac{\pi u}{\lambda_{e \tau}(u)}\Re B_0^{e \tau}(u) - \Im \bar C_0^{e \tau}(u)\right)\Bigg\}. 
\end{align}\\
$s_x$-component:
\begin{align}
    \mathcal{M}_{e\tau}^{(x)} &= \frac{2}{\gamma_\tau}\pi\alpha^3 \lambda s_x \sin\theta \Bigg\{\frac{(t-u)(t+u-4m_\tau^2)}{4t u}\log\frac{m_e^2}{m_\tau^2} + \frac{s m_e^2 (u-t) \Re C_0^{e}(s)}{t u-\left(m_e^2-m_{\tau }^2\right){}^2} \nonumber\\
   &-\frac{t}{2}\frac{\Re B_0^{e \tau}(t)}{\lambda_{e \tau}(t)}\notag\\
   &\quad\times\bigg[\frac{(m_e^2-m_\tau^2+t)(4m_\tau^2+t-u)}{t}
   -2 \left(\frac{s m_e^2(m_\tau^2-m_e^2+t)}{(m_e^2 - m_\tau^2)^2-tu} + 2(m_\tau^2-t)\right) \log\frac{m_{\tau }^2}{s} \bigg] \nonumber\\
   &+\frac{u}{2}\frac{\Re B_0^{e \tau}(u)}{\lambda_{e \tau}(u)} \notag\\
   &\quad\times\bigg[\frac{(m_e^2-m_\tau^2+u)(4m_\tau^2-t+u)}{u}-2 \left(\frac{s m_e^2(m_\tau^2-m_e^2+u)}{(m_e^2 - m_\tau^2)^2-tu} + 2(m_\tau^2-u)\right) \log\frac{m_{\tau }^2}{s} \bigg]\notag\\
 &+\left(\frac{s m_e^2(m_e^2-m_\tau^2-t)}{(m_e^2-m_\tau^2)^2-tu}+2(m_\tau^2-t)\right) \bigg[\Re \bar C_0^{e \tau}(t) + \frac{t}{2}\frac{\Re B_0^{e \tau}(t)}{\lambda_{e \tau}(t)}\log\frac{m_\tau^2}{m_e^2} \bigg]  \nonumber\\
    &-\left(\frac{s m_e^2(m_e^2-m_\tau^2-u)}{(m_e^2-m_\tau^2)^2-tu}+2(m_\tau^2-u)\right) \bigg[\Re \bar C_0^{e \tau}(u) + \frac{u}{2}\frac{\Re B_0^{e \tau}(u)}{\lambda_{e \tau}(u)}\log\frac{m_\tau^2}{m_e^2} \bigg]   
   \Bigg\} \nonumber\\
   &-\frac{8}{\gamma_\tau}\pi\alpha^3 m_e^2 \lambda s^+_x \sin\theta\notag\\
   &\quad\times\bigg[\Re \bar C_0^{e \tau}(u) - \frac{t \Re B_0^{e \tau}(t)}{\lambda_{e \tau}(t)} \left(1 + \log\frac{m_\tau^2}{s}\right) - \frac{u \Re B_0^{e \tau}(u)}{\lambda_{e \tau}(u)}\left(1-\log\frac{m_\tau}{m_e}\right) \bigg] \nonumber\\
   &+\frac{8}{\gamma_\tau}\pi\alpha^3 m_e^2 \lambda s^-_x \sin\theta \notag\\
   &\quad\times\bigg[\Re \bar C_0^{e \tau}(t) -\frac{t \Re B_0^{e \tau}(t)}{\lambda_{e \tau}(t)}\left(1-\log\frac{m_\tau}{m_e}\right)  - \frac{u \Re B_0^{e \tau}(u)}{\lambda_{e \tau}(u)} \left(1+\log\frac{m_\tau^2}{s}\right)\bigg]. 
\end{align}\\
$s_z$-component:
\begin{align}
    \mathcal{M}_{e\tau}^{(z)} &= \frac{4\pi\alpha^3 \lambda s_z }{s^2\beta_e \beta_\tau}\Bigg\{s^2\log\frac{m_e m_\tau}{s} + \frac{s\big(m_\tau^2-m_e^2\big)\big((m_\tau^2+m_e^2)(t+u)-t^2-u^2\big)}{tu} \log\frac{m_e}{m_\tau} \nonumber\\
    &+2s^2m_e^2 \Re C_0^{e}(s) + 2s^2m_\tau^2 \Re C_0^{\tau}(s) +\left(m_e^2+m_\tau^2-t\right)\left[s(t-u)-8m_e^2m_\tau^2\right] \Re \bar C_0^{e \tau}(t) \nonumber\\
    &-\left(m_e^2+m_\tau^2-u\right)\left[s(t-u)+8m_e^2m_\tau^2\right] \Re \bar C_0^{e \tau}(u) \nonumber\\
    &+\frac{u\Re B_0^{e \tau}(u)}{\lambda_{e \tau}(u)}\Bigg[\left(m_e^2+m_{\tau }^2-u\right)s(u-t) \log\frac{m_{\tau }^3}{s m_e} \nonumber\\[0.1cm]
    & -8m_e^2m_\tau^2 \left(m_e^2+m_{\tau }^2-u\right) \log\frac{m_{\tau }}{m_e} +\frac{s\left(m_e^2+m_\tau^2-t\right)\left((m_e^2-m_\tau^2)^2-u^2\right)}{u}\Bigg] \nonumber\\
    &+\frac{t\Re B_0^{e \tau}(t)}{\lambda_{e \tau}(t)}\Bigg[\left(m_e^2 + m_\tau^2 - t\right)s(t-u) \log\frac{m_{\tau }^3}{s m_e}\nonumber\\[0.1cm]
    & -8m_e^2m_\tau^2 \left(m_e^2+m_\tau^2-t\right) \log\frac{m_{\tau }}{m_e} +\frac{s\left(m_e^2+m_\tau^2-u\right)\left((m_e^2-m_\tau^2)^2-t^2\right)}{t} \Bigg]\Bigg\} \nonumber\\
    &+\frac{32 \pi \alpha^3 m_e^2m_\tau^2 \lambda s^-_z(t-u)}{s^2 \beta_e \beta_\tau}\Bigg\{\Re \bar C_0^{e \tau}(u) +\frac{t\Re B_0^{e \tau}(t)}{\lambda_{e \tau}(t)}\Bigg( \frac{m_e^2 + m_\tau^2 - t}{t-u}\log\frac{m_{\tau }^2}{s} - 1\Bigg) \nonumber\\
    & +\frac{u\Re B_0^{e \tau}(u)}{\lambda_{e \tau}(u)}\Bigg(\frac{m_e^2 + m_\tau^2 - t}{t-u} \log\frac{m_{\tau }^2}{s} +\log\frac{m_{\tau }}{m_e} - 1\Bigg)\Bigg\}    \nonumber\\
    &-\frac{32 \pi \alpha^3 m_e^2m_\tau^2 \lambda s^+_z(t-u)}{s^2 \beta_e \beta_\tau}\Bigg\{\Re \bar C_0^{e \tau}(t) +\frac{u\Re B_0^{e \tau}(u)}{\lambda_{e \tau}(u)}\Bigg(\frac{m_e^2 + m_\tau^2 - u}{u-t}\log\frac{m_{\tau }^2}{s}  -1\Bigg) \nonumber\\
    & +\frac{t\Re B_0^{e \tau}(t)}{\lambda_{e \tau}(t)}\Bigg(\frac{m_e^2 + m_\tau^2 - u}{u-t} \log\frac{m_{\tau }^2}{s} + \log\frac{m_{\tau }}{m_e} -1\Bigg)\Bigg\}.  
\end{align}\\
Spin-independent part:
\begin{align}
    \mathcal{M}_{e\tau}^{(0)} &= \frac{4\pi\alpha^3}{s}\Bigg\{(t-u)\bigg[\log\frac{s}{m_e m_\tau} + \frac{1}{\beta_e^2}\log\frac{m_e^2}{s} + \frac{1}{\beta_\tau^2}\log\frac{m_\tau^2}{s} \nonumber\\ 
    & + \frac{(m_\tau^2-m_e^2)(m_\tau^2+m_e^2+s)}{2tu}\log\frac{m_\tau^2}{m_e^2}\bigg] \nonumber\\
    &+\frac{(u-t)}{s}\bigg(\frac{\left(-8 s m_e^2+8 m_e^4+s^2\right)}{\beta_e^2} \Re C_0^{e}(s) + \frac{\left(-8 s m_{\tau }^2+8 m_{\tau}^4+s^2\right)}{\beta_\tau^2}\Re C_0^{\tau}(s)\bigg) \nonumber\\
    &+(s+2t) \left(m_e^2+m_{\tau }^2-t\right) \Bigg(\Re \bar C_0^{e \tau}(t) \nonumber\\
    & +\frac{t \Re B_0^{e \tau}(t)}{\lambda_{e \tau}(t)}\left[\frac{2s}{s+2t} + \log\frac{m_\tau}{m_e} + \left(1 + \frac{4\lambda_{e \tau}(t)+16m_e^2m_\tau^2}{s(s+2t)}\right)\log\frac{m_\tau^2}{s}\right]\Bigg) \nonumber\\[0.1cm]
    &-(s+2u) \left(m_e^2+m_{\tau }^2-u\right) \Bigg(\Re \bar C_0^{e \tau}(u) \nonumber\\[0.1cm]
    & +\frac{u \Re B_0^{e \tau}(u)}{\lambda_{e \tau}(u)} \left[\frac{2s}{s+2u} + \log\frac{m_\tau}{m_e} + \left(1 + \frac{4\lambda_{e \tau}(u)+16m_e^2m_\tau^2}{s(s+2u)}\right)\log\frac{m_\tau^2}{s}\right]\Bigg) \nonumber\\[0.1cm]
    &+\Re B_0^{e \tau}(t)\left(m_e^2 + m_\tau^2 + s - t \right) - \Re B_0^{e \tau}(u)\left(m_e^2 + m_\tau^2 + s - u \right)\Bigg\}. 
\end{align}\\
IR-divergent part:
\begin{align}
    \mathcal{M}_{e\tau}^{(\text{IR})} &= \frac{2\alpha}{\pi\epsilon} \left(\frac{\mu^2}{m_\tau^2}\right)^\epsilon \mathcal{M}_\text{tree}\bigg[\Re B_0^{e \tau}(t)\frac{t(m_e^2+m_\tau^2-t)}{\lambda_{e \tau}(t)} -\Re B_0^{e \tau}(u)\frac{u(m_e^2+m_\tau^2-u)}{\lambda_{e \tau}(u)}\bigg]. 
\end{align}
In the expressions above, we used $s_i = (s_+ + s_-)_i$ and the short form 
\begin{equation}
    \lambda_{e \tau}(x) \equiv \lambda(x, m_e^2, m_\tau^2), 
\end{equation}
for the K\"all\'en function with $x = t,u$.
Moreover, several quantities from Package-X have been abbreviated. We define
\begin{align}
    B_0^{\ell}(s) &\equiv \texttt{DiscB}[s,m_{\ell},m_{\ell}],\qquad 
B_0^{e\tau}(x) \equiv \texttt{DiscB}[x,m_e,m_\tau],\notag\\
\texttt{DiscB}[x,m_e,m_\tau]&=\frac{\lambda^{1/2}_{e\tau}(x)}{x}\log x_s,\qquad 
x_s=-\frac{x-(m_e-m_\tau)^2-\lambda^{1/2}_{e\tau}(x)}{x-(m_e-m_\tau)^2+\lambda^{1/2}_{e\tau}(x)},
\end{align}
where the \texttt{DiscB} function represents the discontinuous part of the scalar one-loop Passarino--Veltman function $B_0$. 
Furthermore, we have defined
\begin{align}
    \bar{C}_0^{\ell}(s) &\equiv \texttt{ScalarC0IR6}[s,m_{\ell},m_{\ell}], \qquad \bar{C}_0^{e\tau}(x) \equiv \texttt{ScalarC0IR6}[x,m_e,m_\tau],\notag\\
\texttt{ScalarC0IR6}[x,m_e,m_\tau]&=\frac{x_s}{m_e m_\tau(1-x_s^2)}\bigg[2\log x_s\log(1-x_s^2)-\frac{1}{2}\log^2 x_s+\frac{1}{2}\log^2\frac{m_e}{m_\tau}\notag\\
&-\frac{\pi^2}{6}+\text{Li}_2\big(x_s^2\big)+\text{Li}_2\Big(1-x_s\frac{m_e}{m_\tau}\Big)+\text{Li}_2\Big(1-x_s\frac{m_\tau}{m_e}\Big)\bigg],    
\end{align}
which corresponds to the finite part of the Ellis--Zanderighi IR divergent triangle 6 integral~\cite{Ellis:2007qk}. Lastly, we have
\begin{align}
    C_0^{\ell}(s) &\equiv \texttt{ScalarC0}[s,m_{\ell}^2,m_{\ell}^2,0,0,m_{\ell}]
    =\frac{4\pi^2+3\log^2\frac{\beta_\ell-1}{\beta_\ell+1}+12\text{Li}_2\Big(\frac{\beta_\ell-1}{\beta_\ell+1}\Big)}{6\beta_\ell s},
\end{align}
where \texttt{ScalarC0} is the scalar Passarino--Veltman three-point function.

\section{Results of the angular distributions in the center-of-mass frame}
\label{app:CMSresults}

\begin{figure}[H]
    \centering
    \includegraphics[width=0.8\linewidth]{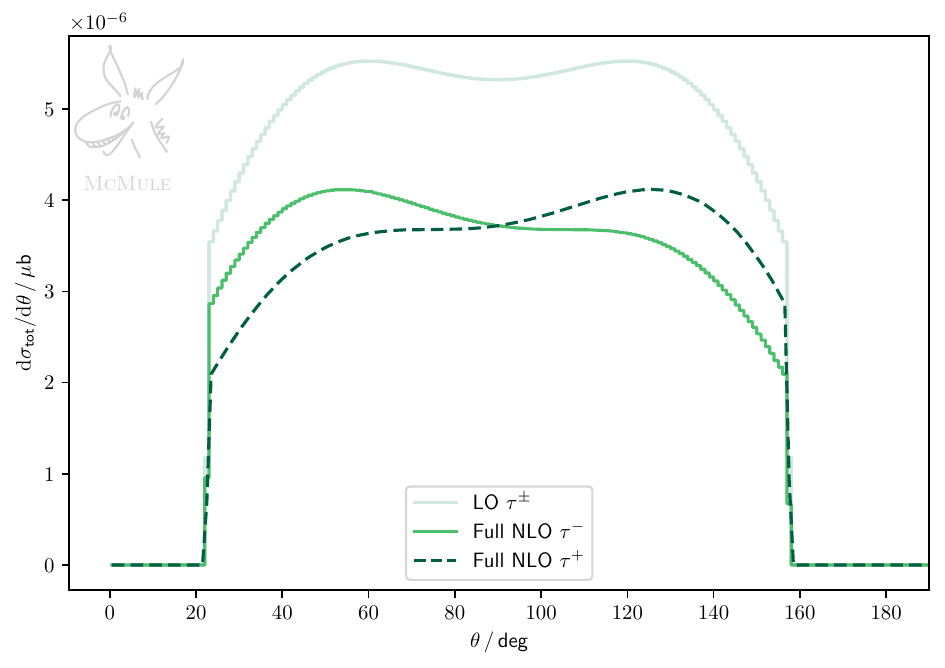}
    \caption{Numerical results of the differential cross section in the CM frame obtained with \mcmule{} including cuts.}
    \label{fig:totcrossCMS}
\end{figure}

\begin{figure}[H]
    \centering
    \includegraphics[width=0.8\linewidth]{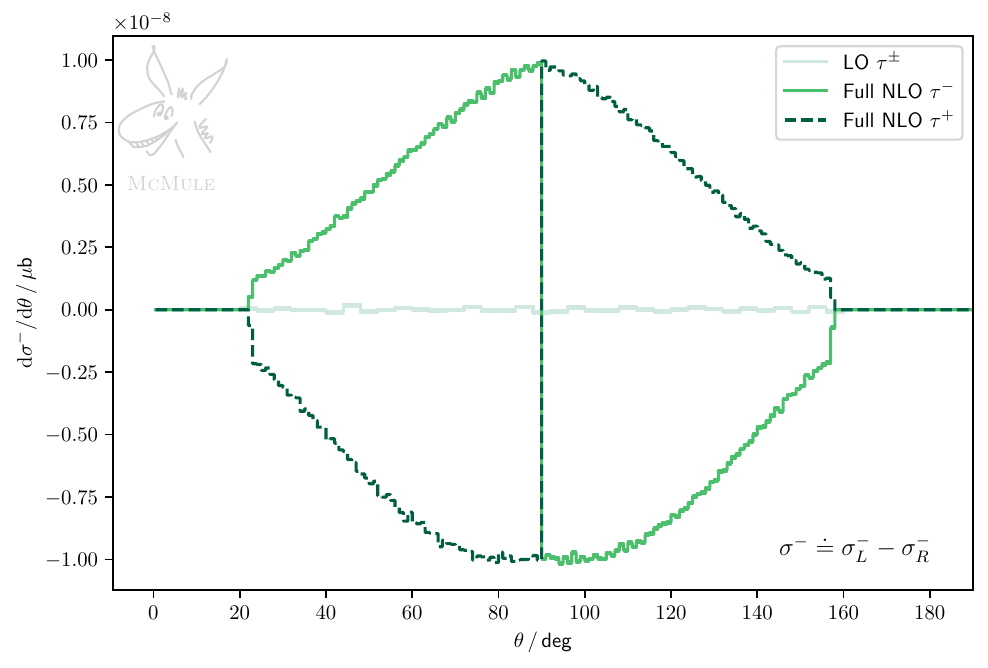}
    \caption{Numerical results of the differential angular distribution of $\A_N^- \times \sigma_\text{tot}$ obtained with \mcmule{} including cuts.}
    \label{fig:ANCMS}
\end{figure}

\begin{figure}[H]
    \centering
    \includegraphics[width=0.8\linewidth]{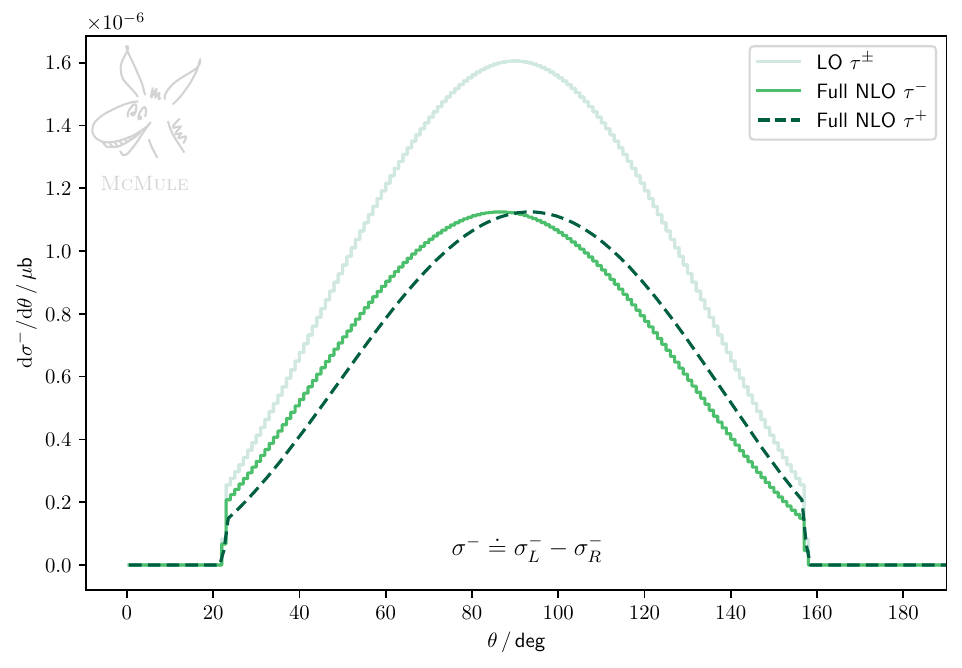}
    \caption{Numerical results of the differential angular distribution of $\A_T^- \times \sigma_\text{tot}$ obtained with \mcmule{} including cuts.}
    \label{fig:ATCMS}
\end{figure}

\begin{figure}[H]
    \centering
    \includegraphics[width=0.8\linewidth]{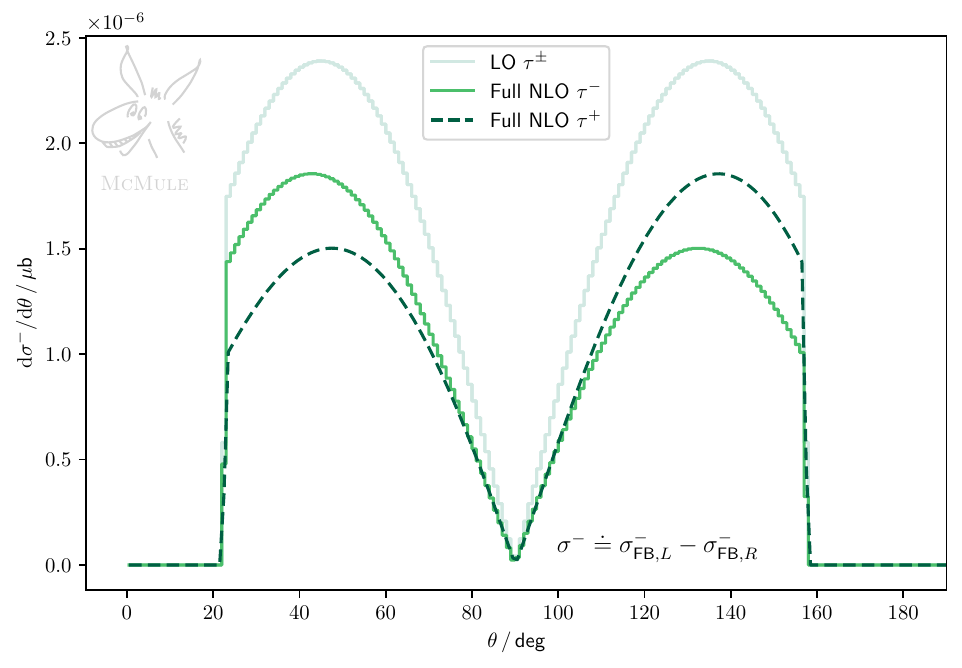}
    \caption{Numerical results of the differential angular distribution of $\A_L^- \times \sigma_\text{tot}$ obtained with \mcmule{} including cuts.}
    \label{fig:ALCMS}
\end{figure}

\bibliographystyle{apsrev4-1_mod_2}
\bibliography{biblio.bib}
	
\end{document}